\documentclass[conference]{IEEEtran}
\IEEEoverridecommandlockouts
\usepackage[letterpaper, top=0.75in, bottom=1.05in, left=0.63in, right=0.63in]{geometry}
\usepackage{cite}
\usepackage{amsmath,amssymb,amsfonts}
\usepackage{algorithmic}
\usepackage{graphicx}
\usepackage{textcomp}
\usepackage{xcolor}
\usepackage{xcolor}
\usepackage{adjustbox}
\usepackage{tikz}
\usepackage{tikz}
\usepackage{multirow}
\def\BibTeX{{\rm B\kern-.05em{\sc i\kern-.025em b}\kern-.08em
    T\kern-.1667em\lower.7ex\hbox{E}\kern-.125emX}}

\usepackage[compatibility=false]{caption}
\DeclareCaptionFont{quackfont}{\fontfamily{ptm}\fontsize{7.5pt}{9pt}\selectfont}
\usepackage[labelformat=simple,font=quackfont]{subcaption}

\DeclareUnicodeCharacter{2212}{\ensuremath{-}}

\makeatother

\begin{document}

\title{A Deep Multi-Task Learning Approach to Impulsive Noise Parameter Estimation\\
%\thanks{This work was supported by the 5G Positioning, Sensing and Security Functions (5G-PSS) through Business Finland under Grant 6868/31/2021 and also is part of the DIOR project that has received funding from the European Union’s MSCA RISE programme under grant agreement No. 10100828.}
}
\author{Abdullahi~Mohammad,~\IEEEmembership{Member,~IEEE,}
    Bdah~Eya,~\IEEEmembership{Student,~IEEE} and
     Bassant~Selim,~\IEEEmembership{Member,~IEEE,}\\
     \IEEEauthorblockA{\IEEEauthorrefmark{1}Department of System Engineering\\
     École de technologie supérieure - ETS Montréal, Quebec, Canada\\.}\\[-4.ex]
e-mail: abdullahi.mohammad@etmtl.ca; eya.bdah.1@ens.etsmtl.ca; bassant.selim@etmtl.ca
}

\maketitle

\begin{abstract}
Impulsive noise poses a significant challenge to the reliability of wireless communication systems, necessitating accurate estimation of its statistical parameters for effective mitigation. This paper introduces a multitask learning (MTL) framework based on a CNN–LSTM architecture enhanced with an attention mechanism for the joint estimation of impulsive noise parameters. The proposed model leverages a unified weighted-loss function to enable simultaneous learning of multiple parameters within a shared representation space, improving learning efficiency and generalization across related tasks. Experimental results show that the proposed MTL framework achieves stable convergence, faster training, and enhanced scalability with modest computational overhead. Benchmarking against conventional single-task learning (STL) models confirms its favorable complexity–performance trade-off and significant memory savings, indicating the effectiveness of the MTL approach for real-time impulsive noise parameter estimation in wireless systems.
\end{abstract}

\begin{IEEEkeywords}
Convolutional neural networks, Impulsive noise
parameter estimation, Long short-term memory networks, Multitask Learning, Single Task Learning.
\end{IEEEkeywords}

\section{Introduction}
\IEEEPARstart{T}{he} next generation of wireless communication systems, six-th generation (6G), and beyond, is envisioned to provide unprecedented levels of reliability, throughput, and low-latency performance, supporting a wide range of applications, including massive machine-type communications, autonomous systems, and integrated sensing and communication (ISAC) \cite{chowdhury20206g,kabir2024complete}. However, as wireless networks become increasingly complex and spectrum-dense, their susceptibility to various forms of interference grows significantly. Among these, impulsive noise is particularly challenging due to its abrupt bursts, high energy, and non-Gaussian nature, which can seriously degrade the quality of service (QoS) and compromise the reliability of the communication system \cite{pighi2009fundamental}. Moreover, impulsive noise is commonly found in urban settings, high-voltage industrial facilities, and indoor environments. Typical sources of this impediment include car ignition systems, electrical power lines, various electronic gadgets, home appliances, medical devices, and industrial machinery \cite{blackard1993measurements,blankenship1998characteristics,sanchez2007interference}. High variations in amplitude characterize its random and unpredictable short-duration spikes, leading to elevated bit error rates (BER), reduced energy efficiency, and even link failures in critical scenarios \cite{alam2019novel}. Traditional communication methods, which often assume Gaussian noise, are ineffective in accurately modeling and handling impulsive interference \cite{jia2015impulsive}. This reveals a significant challenge in making future wireless systems resilient in unpredictable environments.

The detrimental impact of impulsive noise on communication system performance has driven extensive research into receiver-side mitigation strategies based on advanced signal processing. However, their effectiveness strongly depends on the noise type and its parameters \cite{bai2020fifty}. Classical Gaussian-based approaches perform poorly in impulsive environments, motivating the use of more suitable statistical models such as Middleton’s Class-A and the symmetric $\alpha$-stable (S$\alpha$S) distributions, which better characterize impulsive interference. For example, in studies on Middleton Class-A noise \cite{saaifan2012decision,jiang2007bayesian}, Markov Chain Monte Carlo methods were used for parameter estimation, though at the cost of high computational complexity due to the large number of iterations required for convergence. Similarly, Sacuto \textit{et al}. \cite{sacuto2013fuzzy} used fuzzy C-means clustering to estimate partitioned Markov chain impulse noise parameters in power line communications (PLC), limiting their study to three impulsive noise states. Although such works advanced the field, they remain focused on PLC systems under Gaussian channel assumptions. Machine learning (ML) provides a promising alternative given its strength in handling complex wireless communication tasks \cite{mohammad2020complexity}, including resource allocation, beamforming, modulation classification, and signal detection. Within this context, \cite{selim2020effect} proposed a deep neural network (DNN) to classify memoryless Middleton Class-A noise parameters in PLC, while \cite{isac2021impulsive} extended this by combining a convolutional neural network (CNN) and a long short-term memory (LSTM) network to estimate impulsive noise parameters with memory over fading channels. Their results showed that the CNN-LSTM model effectively captures the statistical behavior of impulsive noise from limited received symbols, enabling real-time mitigation. However, the common assumption of memoryless noise limits realism, as noise typically exhibits temporal dependencies \cite{courjault2020robust}. Despite these advances, most ML-based impulsive noise estimators still focus on single-task learning (STL). However, many physical-layer signal processing tasks are interdependent \cite{mohammad2023memory}. They can be better optimized using a multitask learning (MTL) framework that learns related objectives jointly. Despite MTL’s proven success in computer vision, its use in wireless communication remains limited. To address this, we propose an MTL framework for the joint estimation of impulsive noise parameters, which reduces the computational cost of multiple STL models and improves estimation accuracy for more efficient impulsive noise mitigation. The proposed MTL demonstrates superior training stability and consistency, highlighting its strength in learning correlated noise parameters.

\section{System Model Description}
Consider a single-antenna point-to-point wireless communication system. The transmitter transmits an M-ary quadrature amplitude modulation (M-QAM) over a fading channel that is subject to impulsive interference. Therefore, the corresponding complex baseband received signal at the i-th time slot can be expressed as
\begin{equation}\label{received_signal}
y_{i} = h_{i}x_{i} + n_{i},
\end{equation}
where $x_{i}$ denotes the transmitted M-QAM symbol, $h_{i}\sim \mathcal{N}\left( 0,1 \right)$ represents the Rayleigh fading channel coefficient, and $n_{i}$ accounts for the additive noise, which is modeled using a two-state Markov–Gaussian distribution, as in \cite{ndo2013markov}.

Several models characterize impulsive noise \cite{ndo2013markov,shongwe2015study,rouissi2021simulation}, with the Two-State Markov–Gaussian model effectively capturing its temporal correlation ~\cite{ndo2013markov}. Thus, it is adopted in this work. The noise sample is expressed as~\cite{shongwe2015study} 
\begin{equation} n_{i} = bv_{i} + (1-b)w_{i}, \end{equation}
where \( b \sim \text{Bernoulli}(p) \) represents the probability of impulsive noise during a symbol duration. The impulsive component is modeled by \( v_{i} \sim \mathcal{N}(0, \sigma_{v}^{2}) \), and the background Gaussian noise by \( w_{i} \sim \mathcal{N}(0, \sigma_{w}^{2}) \). When \( b = 0 \), only background noise is present; when \( b = 1 \), impulsive noise occurs, i.e., \( \mathcal{P}(b=1) = p \). The impulse-to-Gaussian noise power ratio is defined as: \begin{equation}\label{power_ratio}
 R_{vw} = \frac{\sigma_{v}^{2}}{\sigma_{w}^{2}}. 
\end{equation}

\subsubsection{Two-State Markov–Gaussian Noise Model}
In this model, the noise process alternates between two Gaussian states: a “good ($s_{G}$)” state with low variance corresponding to background noise, and a “bad ($s_{B}$)” state with high variance corresponding to impulsive noise. A first-order Markov chain governs the switching between these states in the $i$-th symbol duration, given as \cite{ndo2013markov}
\begin{equation}
 p(n_{i} \mid s_{i}) \sim \mathcal{N}(0, \sigma_{s_{i}}^{2}),
\end{equation}
where $s_{i}\in \left\{ s_{G},s_{B} \right\}$ denotes the hidden Markov state at time $i$-th symbol duration. The transition probabilities of the underlying Markov chain capture the burstiness of the impulsive events. Without loss of generality, the states of the noise being good and bad are respectively modeled as a first-order Markov process with corresponding transition probabilities, as
\begin{equation}\label{p_g}
p_{G}=
\mathcal{P}(s_{i}=s_{G})=\frac{p_{BG}}{p_{GB}+p_{BG}},
\end{equation}
\begin{equation}\label{p_b}
 p_{B}=
\mathcal{P}(s_{i}=s_{B})=\frac{p_{GB}}{p_{GB}+p_{BG}}.
\end{equation}
Accordingly, the memory associated with the impulsive channel is defined by the inverse sum of the states' transition probabilities, defined as follows
\begin{equation}\label{G}
   \Gamma=\frac{1}{p_{GB}+p_{BG}},
\end{equation}
where $p_{GB}$ and $p_{GB}$ denote the transition probability from state $G$ to state $B$ and vice versa, respectively.

\begin{figure}[t]
    \centering
    \includegraphics[width=3.4in,height=1.8in]{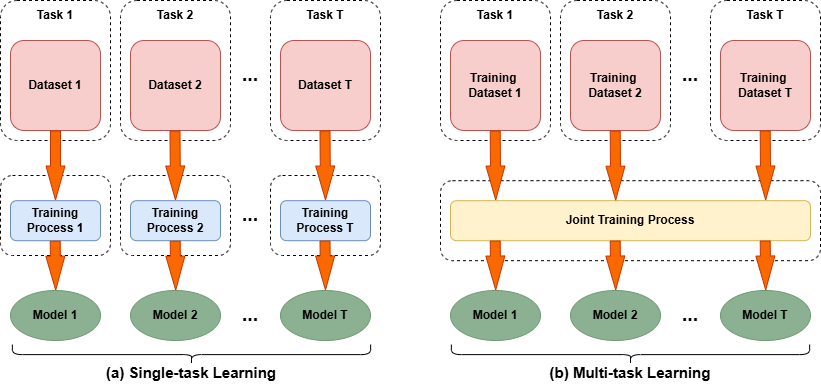}
    \caption{Comparison between STL and MTL workflow}
    \label{fig:STL_MTL_models}
\end{figure}

\section{Proposed MTL Framework for parameter estimation approach}
This section introduces an MTL framework for jointly estimating key parameters that characterize impulsive noise in wireless channels. Specifically, the framework targets the probability of impulsive occurrence, $p$, the impulsive-to-Gaussian power ratio,  $R $, and the channel memory length,  $\Gamma$, all of which are inferred directly from a sequence of received symbols. Since these parameters assume continuous values, the problem formulation naturally corresponds to a regression setting. The MTL enables concurrent learning of all parameters within a unified architecture. To highlight this, Fig. \ref{fig:STL_MTL_models}a illustrates the STL scenario, where each task is trained in isolation, whereas Fig. \ref{fig:STL_MTL_models}b depicts the MTL paradigm, where multiple tasks are optimized jointly through a shared model structure.

\begin{table}[t]
\caption{MTL Architecture: Layers and Parameters}
\centering
\scriptsize
\setlength{\tabcolsep}{2.2pt}
\begin{tabular}{|p{1.6cm}|p{2.4cm}|p{2.8cm}|r|}
\hline
\textbf{Block} & \textbf{Layer} & \textbf{Details} & \textbf{Params} \\
\hline
\multirow{8}{*}{Shared Trunk} 
  & Conv1D ($1\!\rightarrow\!32$) & Kernel=8, Stride=2 & 288 \\
  & BN (32) & BatchNorm1d & 64 \\
  & Conv1D ($32\!\rightarrow\!64$) & Kernel=4, Stride=2 & 8256 \\
  & BN (64) & BatchNorm1d & 128 \\
  & Conv1D ($64\!\rightarrow\!128$) & Kernel=4, Stride=2 & 32896 \\
  & BN (128) & BatchNorm1d & 256 \\
  & Attention & Mean + Softmax & 0 \\
  & LSTM ($128\!\rightarrow\!64$) & 1 layers, batch\_first & 74496 \\
  & LSTM ($64\!\rightarrow\!64$) & 2 layers, batch\_second & 2080 \\
\hline
Shared Feature & Extract & $\mathbf{x}[:, -1, :]$ & -- \\
\hline
Head $p$ & FC($64\!\rightarrow\!1$) & Regression & 65 \\
\hline
\multirow{3}{*}{Head $R$} 
  & FC($64\!\rightarrow\!64$) & ReLU & 4160 \\
  & FC($64\!\rightarrow\!32$) & ReLU & 2080 \\
  & FC($32\!\rightarrow\!1$) & Output & 33 \\
\hline
\multirow{4}{*}{Head $\Gamma$} 
  & FC($64\!\rightarrow\!64$)+BN & ReLU, BN(64) & 4288 \\
  & Dropout & $p=0.4$ & 0 \\
  & FC($64\!\rightarrow\!32$)+BN & ReLU, BN(32) & 2144 \\
  & FC($32\!\rightarrow\!C$) & $C=4$ classes & 132 \\
\hline
\textbf{Total} & & & \textbf{183,462} \\
\hline
\end{tabular}
\label{tab:mtl}
\end{table}

\begin{figure*}[t]
    \centering
    \includegraphics[width=6.8in,height=2.8in]{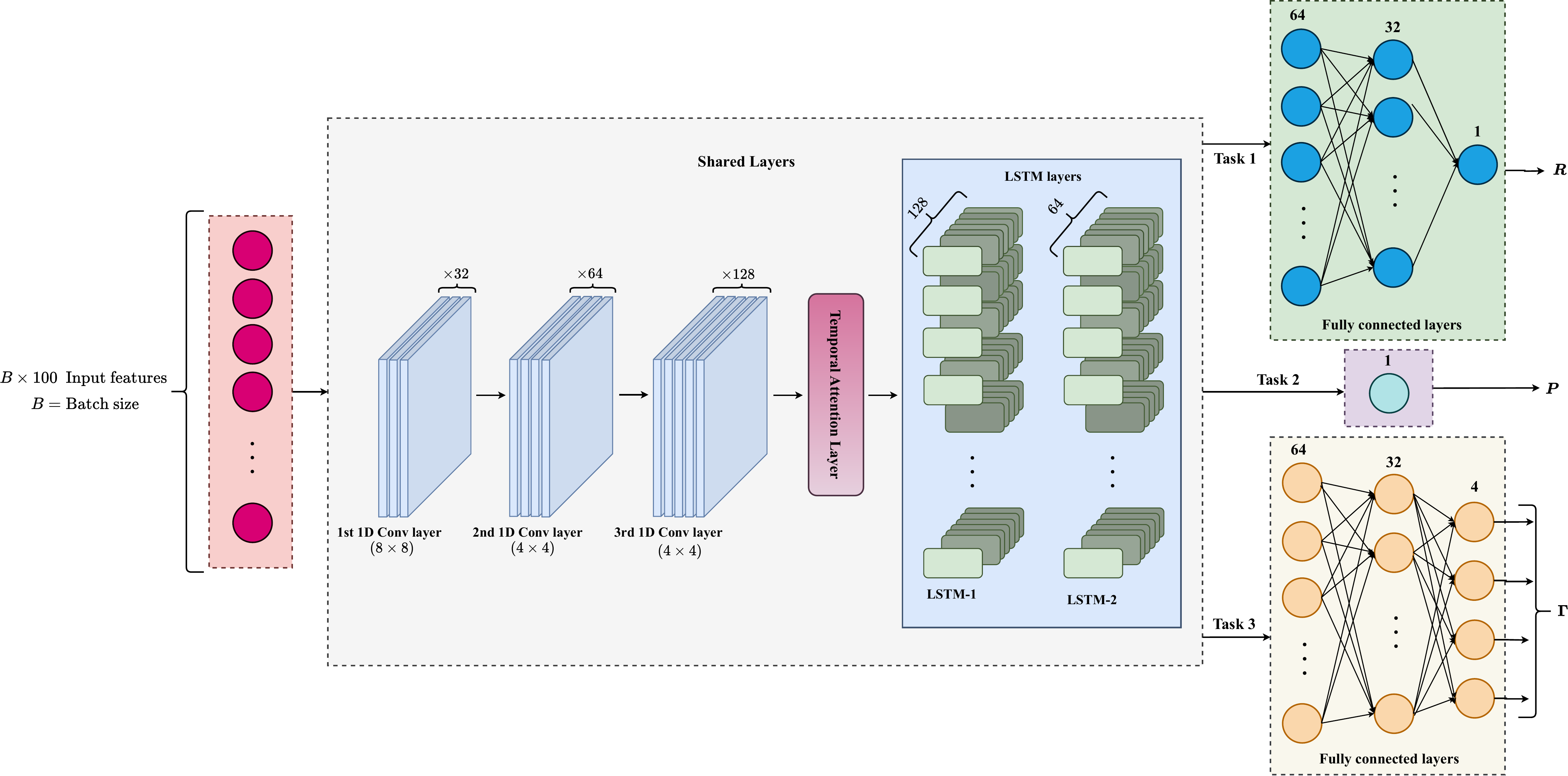}
    \caption{Proposed Multi-tasks Deep CNN-LSTM model}
    \label{fig:MTL_model}
\end{figure*}

\subsection{Proposed MTL Architecture}
The proposed MTL architecture leverages a shared CNN–LSTM backbone layer and a temporal attention module with multiple task-specific heads to enable the joint estimation of impulse noise parameters. The model enables shared representation learning and exploits inter-task correlations. The overall proposed model architecture is depicted in Fig.~\ref{fig:MTL_model} and also summarized in Table \ref{tab:mtl}. The input to the network is a sequence of received baseband symbols of dimension $\mathcal{B} \times 100$, where $\mathcal{B}$ denotes the batch size. These samples contain both Gaussian and impulsive components, which form the raw input features for subsequent processing. The feature extraction stage consists of three one-dimensional convolutional (Conv1D) layers with batch normalization and ReLU activation \cite{mohammad2023learning}:
\begin{itemize}
    \item \textbf{Conv1:} $1 \!\to\! 32$ filters, kernel size $=8$, stride $=2$. This layer captures local dependencies and short impulsive bursts.
    \item \textbf{Conv2:} $32 \!\to\! 64$ filters, kernel size $=4$, stride $=2$. This layer extracts mid-level abstractions that capture correlations between consecutive noise bursts.
    \item \textbf{Conv3:} $64 \!\to\! 128$ filters, kernel size $=4$, stride $=2$. This layer learns high-dimensional representations to distinguish between impulsive and Gaussian features.
\end{itemize}

A temporal attention mechanism is applied to adaptively weight features across time. This mechanism guarantees symbol intervals dominated by impulsive noise while suppressing Gaussian-like regions, thereby enhancing discriminative power. Two stacked LSTM layers with 64 neurons capture long-range temporal dependencies. Since impulsive noise exhibits bursty and memory-dependent behavior, the LSTM modules provide a packed sequence-level representation of the received symbols. The CNN layers, the temporal attention unit, and the LSTM layers form the shared layer across the three tasks. The output of the final LSTM cell is used as the shared representation for all tasks. From the shared representation, three task-specific fully connected (FC) branches predict the parameters:
\begin{itemize}
    \item \textbf{Head for $R$:} A two-layer FC branch $64 \!\to\! 32 \!\to\! 1$ is used to estimate the impulsive-to-Gaussian power ratio, which requires deeper feature processing due to its dependence on both impulsive and Gaussian components.
    \item \textbf{Head for $p$:} A single FC layer $64 \!\to\! 1$ is employed to estimate the probability of impulsive occurrence, which can be directly inferred from the shared representation.
    \item \textbf{Head for $\Gamma$:} A three-layer FC branch $64 \!\to\! 32 \!\to\! C$ with dropout regularization predicts the channel memory length, modeled as a classification problem with $C$ discrete memory classes.
\end{itemize}

\subsection{Experimental Setup}\label{exper_setting}
The dataset used in this work was synthetically generated from transmitted symbols drawn from a normal distribution to simulate quadrature phase shift keying (QPSK)-modulated signals subjected to impulsive noise, modeled using a Markov–Gaussian process. For each configuration of the noise parameters ($p, R,\Gamma$), a total of 40,000 sequences, each consisting of 100 QPSK symbols, were generated, yielding approximately 1.3 million sequences overall. The parameter values considered were: $p = \{0, 0.1\}$ (impulsive noise occurrence probability), $R = \{1, 10, 100, 1000\}$ (impulsive-to-Gaussian ratio), and $\Gamma = \{1, 10, 50, 100\}$ (channel memory). Each sequence comprised 100 time-domain samples serving as features, while the corresponding triplet $(p, R, \Gamma)$ was used as the target label. Consequently, the dataset represents the received symbols derived from equations (\ref{received_signal}), (\ref{power_ratio}), (\ref{p_b}), (\ref{p_g}), and (\ref{G}), under a training signal-to-noise-ratio (SNR) of 15 dB. Preprocessing steps were applied to improve model performance, stability, and training convergence. Specifically, features were normalized to a common scale, small feature values were thresholded to zero, a logarithmic transformation was applied to $R$, inputs and targets were standardized, and $\Gamma$ was encoded. Finally, the dataset was partitioned into 70\% training, 15\% validation, and 15\% testing subsets, with stratification based on $\Gamma$ to preserve class distributions.

\subsection{Model Training and Testing}
The training strategy employs a joint loss function composed of two mean squared error (MSE) terms for the regression tasks $p$ and $R$, and a label-smoothed cross-entropy (LSCE) term for the classification task $\Gamma$. The underlying rationale of MTL is that the shared feature extractor, comprising convolutional, attention, and LSTM layers, learns a common representation that captures general sequential dependencies useful across all tasks. Task-specific output heads then refine this shared representation for their respective objectives. Such a design improves generalization and enhances robustness, as the complementary tasks provide mutual regularization and help mitigate overfitting. Formally, the training objective minimizes the weighted sum of three task-specific loss functions
\begin{equation}\label{eq:mtl_loss}
\mathcal{L}_{\text{MTL}}(\theta) = \underset{\theta}{\text{min}} \sum_{i=1}^{T} \lambda_i \, \mathcal{L}_i(\theta, \mathcal{D}^{\text{train}}), 
\end{equation}
where $\mathcal{D}^{\text{train}}=\left\{y_{i}\right\}_{i=1}^{N}$ is the input training sample containing the true noise parameters $\{p_i,R_i,\Gamma_i\}$, $T$ is the number of tasks, $N$ is the training sample, $\lambda_i$ are non-negative task-specific weighting coefficients and $\theta$ denotes the trainable parameters of the model. The first two components correspond to regression losses for $R$ and $p$, defined respectively as
%\begin{align}
%\mathcal{L}_{p}(\theta) &=\frac{1}{N}\sum_{i=1}^{N}\left( p_{i}-f_{p}(y_{i},\theta) \right)^{2}, \label{eq:loss_p} \\
%\mathcal{L}_{R}(\theta) &=\frac{1}{N}\sum_{i=1}^{N}\left(R_{i}-f_{R}(y_{i},\theta)\right)^{2}, \label{eq:loss_r}
%\end{align}
\begin{align}
\mathcal{L}_{p}^{\text{NMSE}}(\theta) &= \frac{\sum_{i=1}^{N} \left( p_i - f_p(y_i, \theta) \right)^2}{\sum_{i=1}^{N} \left( p_i - \bar{p} \right)^2}, \\
\mathcal{L}_{R}^{\text{NMSE}}(\theta) &= \frac{\sum_{i=1}^{N} \left( R_i - f_R(y_i, \theta) \right)^2}{\sum_{i=1}^{N} \left( R_i - \bar{R} \right)^2},
\end{align}
where $p_{i}$ and $R_{i}$ denote the true parameter values, and $f_{p}(y_{i}, \theta)$ and $f_{R}(y_{i}, \theta)$ are their corresponding predictions. The terms $\bar{p}$ and $\bar{R}$ represent the mean values of the true parameters. The normalized mean squared error (NMSE) is used to achieve scale-invariant evaluation across tasks. Given that the MTL framework involves predicting impulse noise parameters with varying magnitudes, NMSE ensures a fair performance comparison by normalizing each task’s error with respect to the variance of its targets.

The third component is a classification loss for $\Gamma_{i}$, formulated using LSCE to reduce overconfidence and enhance generalization
\begin{align}
\mathcal{L}_{\Gamma}(\theta) &= -\frac{1}{N}\sum_{i=1}^{N}\sum_{c=1}^{K} q_{i,c}\,\log\!\left(\text{softmax}(f_{\Gamma}(y_{i},\theta))_c\right), \label{eq:loss_gamma}
\end{align}
where $K=4$ is the number of classes, as already defined, $f_{\Gamma}(y_{i},\theta)$ denotes the logit vector for the $i$-th sample, and $q_{i,c}$ is the smoothed target distribution defined as
\begin{equation}
q_{i,c}=
\begin{cases}
1-c, & c = {\Gamma}_{i}, \\
\frac{c}{K-1}, & c \neq {\Gamma}_{i},
\end{cases}
\qquad r=0.1\ (\text{smoothing parameter}). \label{eq:label_smoothing}
\end{equation}
where ${\Gamma}_{i}$ the true class index. Thus, the true class is assigned a probability $0.9$, while the remaining $0.1$ is uniformly distributed among the other classes. Therefore, the overall loss function is
\begin{equation}
\mathcal{L}_\text{MTL}(\theta)=\lambda_{1}\mathcal{L}_{p}^{\text{NMSE}}(\theta)+\lambda_{2}\mathcal{L}_{R}^{\text{NMSE}}(\theta)+\lambda_{3}\mathcal{L}_{\Gamma}(\theta).
\end{equation}

By jointly minimizing \eqref{eq:mtl_loss}, the model concurrently (i) regresses the parameter $p$, (ii) estimates $R$ following logarithmic stabilization, and (iii) classifies $\Gamma$ with enhanced robustness against noise. In this study, equal task weighting ($\lambda_1 = \lambda_2 = \lambda_3$) is initially adopted under the assumption that all tasks contribute equally to the overall optimization objective. Furthermore, asymmetric weighting configurations are also investigated, specifically, $\lambda_1 = 0.7, \lambda_2 = 0.85, \ \text{and}\ \lambda_3 = 1.0$, to assess the impact of task prioritization on joint learning performance. The model was trained for over 50 training epochs. During this process, the model receives batches of input samples that are processed through shared layers to capture generalizable features, followed by task-specific branches that refine these features for each prediction objective. The loss functions from all tasks are computed and combined as a weighted sum to guide parameter updates through backpropagation and gradient descent, implemented using the Adam optimizer with a learning rate of 0.001 and a batch size ($\mathcal{B}=64$). To enhance training stability and generalization, batch normalization, dropout, and early stopping are used to mitigate overfitting.

In the inference phase, the trained MTL model leverages the shared and task-specific representations learned during training to make predictions on unseen data. An input sample passes through the shared feature extraction layers, ensuring consistent representation across tasks, before being directed into each task-specific output head. Each branch produces its respective prediction (e.g., regression values, classification probabilities, or sequence labels), allowing the model to solve multiple objectives simultaneously in a single forward pass. Since no parameter updates occur during inference, the process is computationally efficient and exploits the synergy gained from joint training to achieve improved accuracy and robustness compared to training separate STL models.

\begin{figure*}[t]
    \centering
    \begin{subfigure}[b]{0.472\linewidth}
        \centering
        \includegraphics[width=\linewidth]{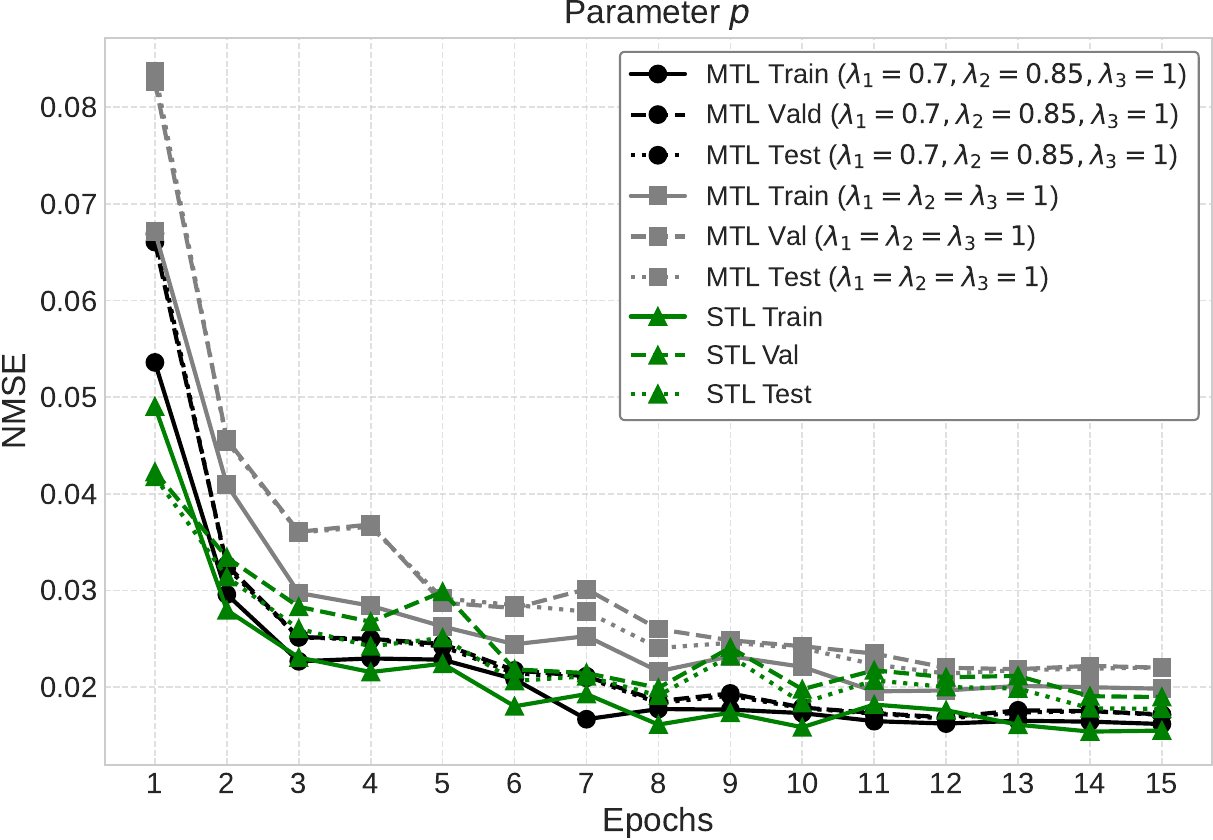}
        \caption{Training, validation and test performance of MTL for estimating $p$}
        \label{fig:MTL1_P}
    \end{subfigure}
    \hfill
    \begin{subfigure}[b]{0.472\linewidth}
        \centering
        \includegraphics[width=\linewidth]{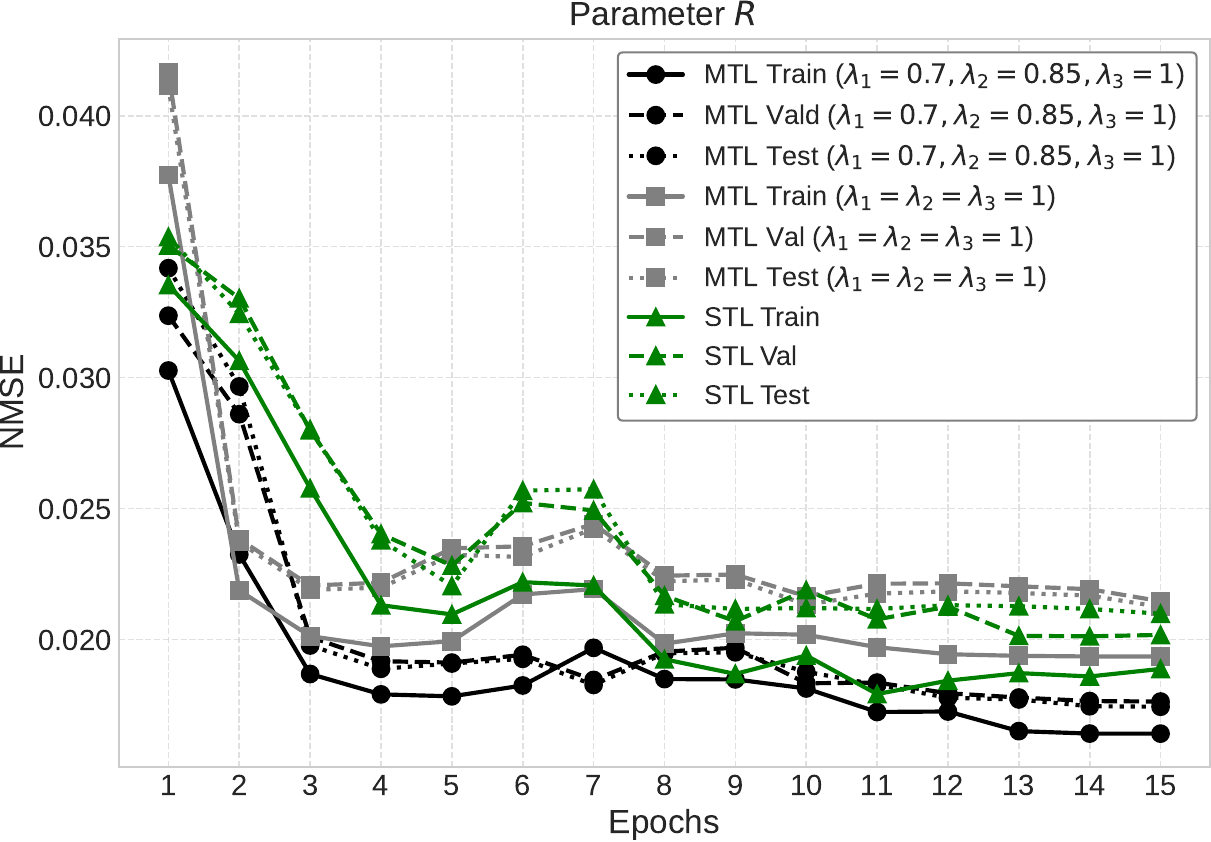}
        \caption{Training, validation and test performance of MTLL for estimating $R$}
        \label{fig:MTL1_R}
    \end{subfigure}
    \caption{Performance of MTL with equal and unequal weighted loss penalties over training epochs.}
    \label{fig:mtl_vs_stl_EW}
\end{figure*}

\begin{figure}[t]
    \centering
    \includegraphics[width=3.25in,height=2.4in]{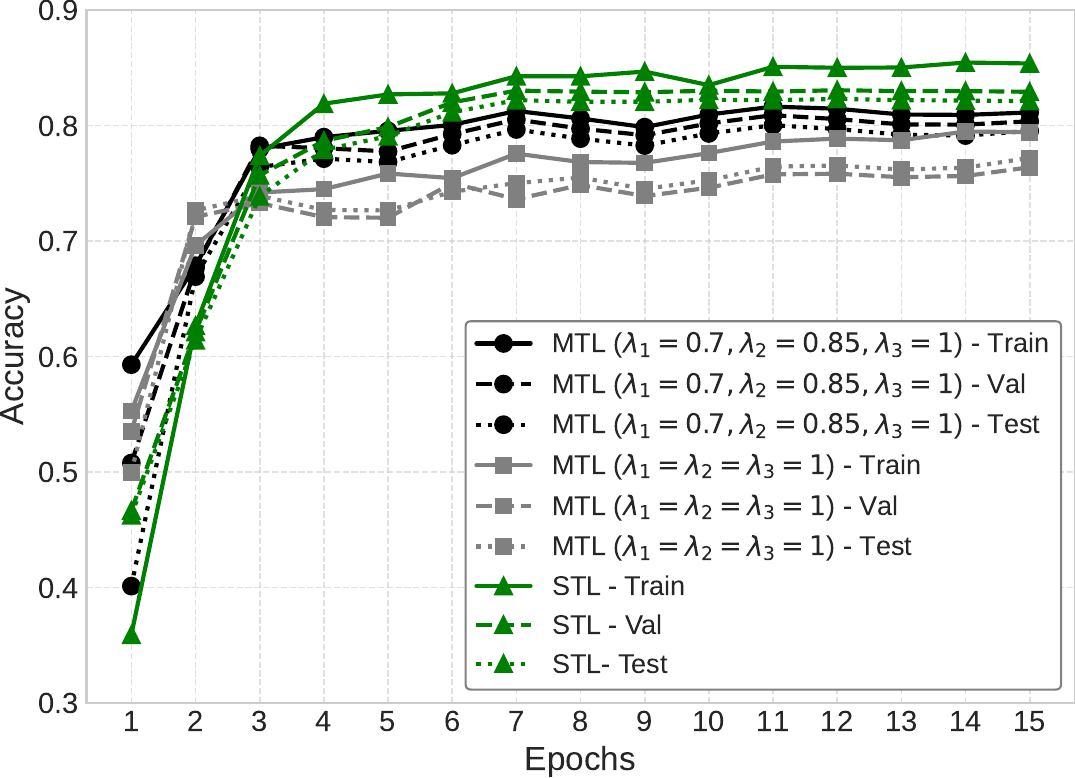}
    \caption{Prediction accuracy of the proposed MTL model and STL for estimating $\Gamma$ parameter.}
    \label{fig:G_pred_accr}
\end{figure}

\begin{figure}[t]
    \centering
    \includegraphics[width=3.25in,height=2.4in]{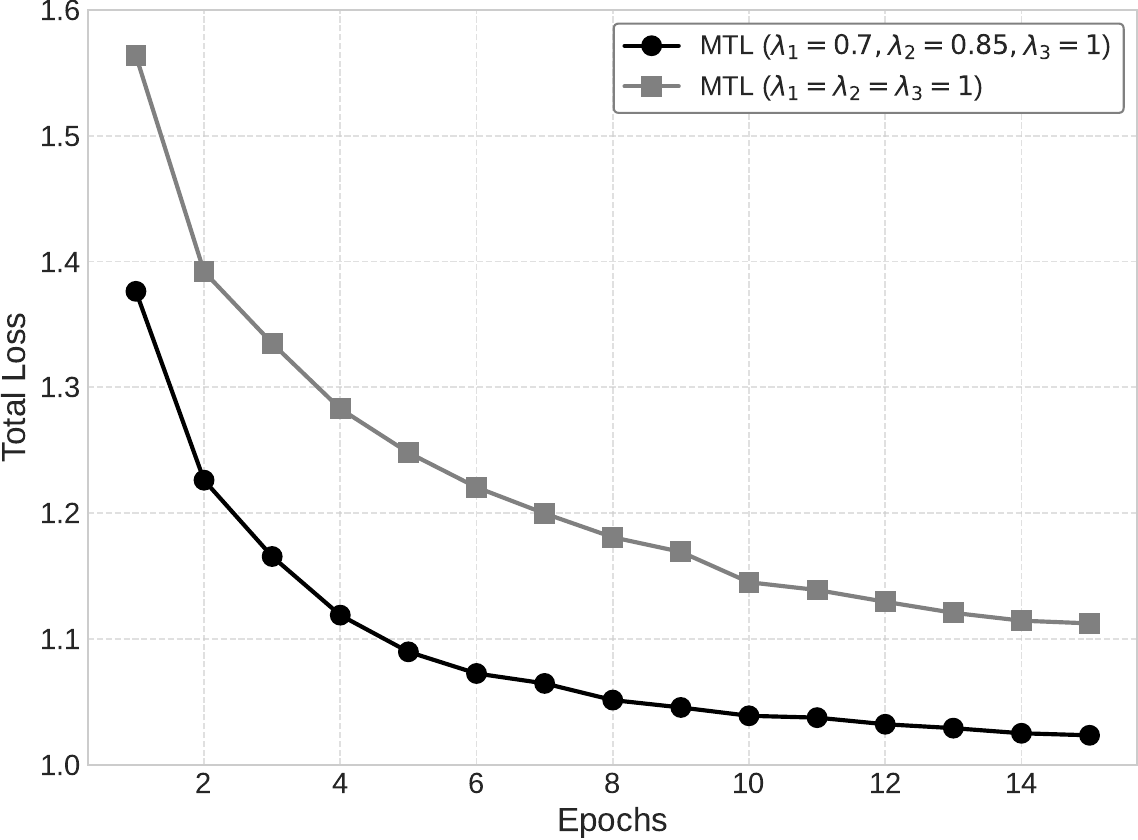}
    \caption{Overall MTL training loss under equal and unequal weighted loss penalty.}
    \label{fig:MTL_Loss}
\end{figure}

\section{Results and Discussions}
To validate the efficacy of the proposed approach, we benchmark its results against those reported in \cite{isac2021impulsive}, where an STL methodology was employed based on the experimental setup outlined in Section \ref{exper_setting}. Fig. \ref{fig:mtl_vs_stl_EW} compares the NMSE performance of the proposed MTL model trained with equal and unequal weighted loss functions for estimating the impulsive noise parameters, $p$ and $R$. Under the equal-weighted configuration ($\lambda_1 = \lambda_2 = \lambda_3 =1$), both models exhibit fast convergence within the first few epochs (Fig. \ref{fig:MTL1_P}). For parameter $p$, STL attains a slightly lower final NMSE than MTL, indicating marginally better precision in single-task optimization. However, the MTL curves are noticeably smoother across training, validation, and testing, indicating more stable convergence and improved generalization due to shared feature representations. In contrast, for parameter $R$, the MTL model performs comparably to STL, particularly in the later epochs, highlighting its robustness in learning correlated parameters. When task-specific weighting is introduced ($\lambda_1 = 0.7, \lambda_2 = 0.85, \lambda_3 = 1$) as shown in Fig. \ref{fig:MTL1_R}, MTL consistently outperforms STL for both parameters, achieving faster convergence and lower NMSE. This shows that adaptive task weighting enhances the efficiency and balance of multitask optimization. While STL occasionally yields marginally higher accuracy on isolated tasks in an equal-weighted scenario, MTL demonstrates superior stability and scalability. 

Unlike the continuous regression modeling of parameters $p$ and $R$, predicting $\Gamma$ is more challenging due to its discrete and sparse distribution. In our experiment, $\Gamma$ takes four distinct values (1, 10, 100, 500), reflecting the memory and temporal structure of impulsive noise. To address this, the task was reformulated as a multi-class classification. As shown in Fig. \ref{fig:G_pred_accr}, both MTL and STL models exhibit a rapid rise in accuracy during early epochs. STL achieves slightly higher accuracy than MTL with unequal-weighted loss, though the latter outperforms the equal-weighted variant. Notably, both MTL configurations show smoother and more stable learning curves across training, validation, and test sets, while STL shows greater fluctuations. This implies that despite STL’s marginally better accuracy, MTL offers superior training stability and consistency, highlighting its strength in learning correlated noise parameters. Fig. \ref{fig:MTL_Loss} illustrates the enhanced robustness of the MTL model trained with a varied-weighted loss, where the combined tasks' loss decreases more rapidly and converges faster across training epochs compared to its equal-weighted counterpart.

\subsection{Computational Complexity Evaluation}
The computational complexity of the proposed MTL framework was evaluated against three STL baselines in terms of the number of learnable parameters, training efficiency, and inference latency, as summarized in Table~\ref{tab:complexity}. It can be observed that despite integrating multiple objectives, the MTL model is 15\% larger than the largest STL-$\Gamma$ model while encapsulating all three tasks. Deploying the three STL models independently would require 240,998 parameters, representing a 31\% increase, thereby confirming the compactness of the MTL architecture. In terms of training cost, the MTL model converges $\sim$47\% faster than the slowest STL model (STL-$p$) and 16\% faster than the fastest (STL-$\Gamma$). The combined STL training time (3617.72 s) is more than four times that of the MTL model, demonstrating a marked reduction in computational overhead and improved temporal efficiency during training. The main trade-off lies in inference latency, where the MTL model is 28\% slower than the average STL model due to simultaneous multitask output computation. However, it remains 56.5\% faster than running all three STL models sequentially. Finally, the MTL framework achieves a favorable complexity–performance balance, providing substantial gains in parameter efficiency and training scalability with a modest inference overhead.
\begin{table}[t]
\caption{Computational Complexity: MTL vs STL}
\label{tab:complexity}
\centering
\begin{tabular}{|l|r|r|r|}
\hline
\textbf{Model} & \textbf{Model Size (KB)} & \textbf{Inference (sec)} & \textbf{Training (sec)} \\
\hline
STL-$p$ & 40,529 & 4.814 & 1582.07 \\
STL-$R$ & 40,529 & 4.816 & 1046.31 \\
STL-$\Gamma$ & 159,940 & 4.990 & 989.34 \\
\hline
\textbf{STL Total} & \textbf{240,998} & \textbf{14.620} & \textbf{3617.72} \\
\hline
MTL & 183,462 & 6.364 & 831.60 \\
\hline
\textbf{Difference} & \textbf{-57,536} & \textbf{-8.256} & \textbf{-2786.12} \\
\textbf{Reduction (\%)} & \textbf{23.9\%} & \textbf{56.5\%} & \textbf{77.0\%} \\
\hline
\end{tabular}
\end{table}

\section{Conclusion}
This paper presented a CNN-LSTM-based MTL framework with an attention mechanism for joint estimation of impulsive noise parameters. Compared to conventional STL techniques, the proposed model achieves smoother convergence, enhanced robustness, and faster training and inference. Despite being only 15\% larger than the largest STL model, it replaces three independent networks that together require 31\% more parameters, reducing overall computational cost. The MTL model trains up to 47\% faster than the slowest STL model and achieves 56.5\% faster inference compared to executing all STL models sequentially, with only a modest 28\% latency increase relative to a single STL model. These results demonstrate that the proposed MTL framework offers an efficient and scalable solution for impulsive noise parameter estimation. %Future work will investigate the effectiveness of the proposed MTL framework in a multi-antenna system configuration, representing a more realistic scenario for next-generation wireless communication systems where interference effects are more pronounced.

\bibliographystyle{IEEEtran}
\bibliography{ref}
\end{document}